\documentclass[]{emulateapj}
\usepackage{graphicx}
\usepackage{amsmath}
\usepackage{xspace}
\usepackage{subfigure}
\usepackage{lineno}
\usepackage{natbib}
\usepackage{comment}
\usepackage{hyperref}

\def\AJ {AJ}
\def\aj {AJ}

\def\pasp {PASP}

\def\apjs {ApJS}
\def\aaps {A\&AS}
\def\aap {A\&A}
\def\apj {ApJ}
\def\apjl {ApJL}
\def\mnras{MNRAS}

\newcommand{\forloop}[5][1]%
{%
\setcounter{#2}{#3}%
\ifthenelse{#4}%
	{%
	#5%
	\addtocounter{#2}{#1}%
	\forloop[#1]{#2}{\value{#2}}{#4}{#5}%
	}%
	{%
	}%
}% 

\newcommand{\ctbd}[1]{}

\newcommand{\lc}{light curve}
\newcommand{\lcs}{light curves}

\newcommand{\rlc}{{\bf RAWLC}}

\newcommand{\epdlc}{{\bf EPDLC}}
\newcommand{\tfalc}{{\bf TFALC}}
\newcommand{\coslc}{{\bf COSLC}}

\newcommand{\rearth}{\ensuremath{R_\oplus}}

\begin{document}
\title{High Precision Photometry for K2 Campaign 1\\ }
\author{C.~X.~Huang\altaffilmark{1}, 
        K.~Penev\altaffilmark{1},
        J.~D.~Hartman\altaffilmark{1}, 
        G.~\'A.~Bakos\altaffilmark{1,2,3}, 
        W.~Bhatti\altaffilmark{1}, 
        I.~Domsa\altaffilmark{1}, 
        M. de Val-Borro\altaffilmark{1}
}
\altaffiltext{1}{Department of Astrophysical Sciences, Princeton 
University, Princeton, NJ 08544; email: chelsea@astro.princeton.edu}
\altaffiltext{2}{Sloan Fellow}
\altaffiltext{3}{Packard Fellow}
\altaffiltext{4}{Hungarian Astronomical Association, Budapest, Hungary}

\begin{abstract}
The two reaction wheel {\em K2} mission promises and has delivered 
new discoveries in the stellar and exoplanet fields. However, 
due to the loss of accurate pointing, it also brings new 
challenges for the data reduction processes. In this paper, 
we describe a new reduction pipeline for extracting high 
precision photometry from the {\em K2} dataset, and present
public light curves for the {\em K2} Campaign 1 target pixel 
dataset. Key to our reduction is the derivation of global 
astrometric solutions from the target stamps, from which accurate 
centroids are passed on for high precision photometry 
extraction. We extract target light curves
for sources from a combined UCAC4 and EPIC catalogue -- this includes
not only primary targets of the {\em K2} campaign 1, but 
also any other stars that happen to fall on the pixel stamps. 
We provide the raw light curves, and the products of various 
detrending processes aimed at
removing different types of systematics. Our astrometric solutions
achieve a median residual of $\sim 0.127^{\prime\prime}$. For 
bright stars, our best 6.5 hour precision for raw \lcs\ is 
$\sim20$ parts per million (ppm). For our detrended \lcs, 
the best 6.5 hour precision achieved is $\sim15$ ppm. We show that our 
detrended \lcs\ have fewer systematic effects (or trends, or red-noise) 
than light curves produced by other 
groups from the same observations. Example light curves of
transiting planets and a Cepheid variable candidate, are also
presented. We make all light curves public, including the 
raw and de-trended photometry, at \url{http://k2.hatsurveys.org}.

\end{abstract}


\section{Introduction}
The {\em Kepler} spacecraft ended its primary mission after the failure 
of two reaction wheels. The {\em K2} mission uses the {\em Kepler} spacecraft
to perform 80-day observations of selected fields in the ecliptic
plane. This brings new opportunities to study transiting 
planets around different stellar populations compared to the 
original {\em Kepler} field, such as clusters of young and pre-main 
sequence stars \citep{Howell:2014}. 

{\em K2} uses the remaining two reaction wheels, and solar
radiation pressure, to maintain close to constant 
pointing of the spacecraft over the 80-day per-field observations.
Currently, observations are performed with 21 modules, each 
module consisting of 4 CCD channels, yielding 76 channels (2 modules 
failed). Due to the limited bandwidth, 
only postage stamps containing proposed targets are downloaded. 
These postage stamps are typically 25$\times$25 pixels 
in size (depending on the brightness of the targets 
and campaigns). These make up only less than 
10$\%$ percent of the entire field of view (FOV). The majority of 
stamps are observed at $\sim$ 30 
minutes cadence. Typically, two Full Field Images (FFIs) 
are downloaded for the beginning and the end of campaign.

However, the two reaction wheel mode also brings in new 
challenges for the data reduction processes. 
The spacecraft pointing is less stable compared to the primary mission, leading to a 
potential decrease in the photometric precision. 
Although the disturbance from the solar pressure is mostly controlled by 
the two reaction wheels and the thruster firing (every 2 days), there is 
still a low frequency motion remaining, resulting in the targets drifting 
across the field of view. The extracted aperture photometry light
curves are dominated by the systematics induced by this drift 
pattern. \citet{VanderburgJohnson:2014} (hereafter VA14) minimized 
this drift systematic by decorrelating the light curves with the motion 
of the spacecraft. They achieved a photometric precision that is within a factor 
of two of the original {\em Kepler} photometry. Various other teams also 
developed their own tools to reduce the {\em K2} data. \citet{Aigrain:2015} 
used aperture photometry and a semi-parametric Gaussian process model 
to extract photometry from the {\em K2} engineering data. 
\citet{Lund:2015} presented 
K2P, a pipeline specifically designed for astrometric analyses.
\citet{Foremanmackey:2015} and \citet{Angus:2015} proposed 
a method to analysis the {\em K2} data without a general detrending process. 

There is, however, room for further improvements. VA14 reduction 
achieved the highest precision among all the past works, but only 
derived photometry for the proposed {\em Kepler} targets (not all 
targets falling on silicon), and are 
also known to have remaining systematic variations affected 
by the spacecraft roll \citep{Angus:2015}. 
\citet{Aigrain:2015} and \citet{Lund:2015} 
derive photometry for all of the targets on silicon, but 
achieved slightly lower precision than VA14, especially for 
the bright stars. 
Here we present a new reduction of the {\em K2} data drawing on 
techniques used in analysing data from ground-based 
surveys \citep[e.g.]{Bakos:2010}.

We approach the {\em K2} pixel file reduction with the following steps: 
1) improved astrometry for source centroiding and 
flux extraction; 2) photometric extraction for {\em all} the 
stars observed on the {\em K2} postage stamps; 
3) removal of first order systematics via a 
modified External Parameter Decorrelation (EPD) procedure 
(broadly similar to VA14); 4) further reduction of the shared systematic 
trends via an implementation of the Trend Filtering Algorithm (TFA) 
and semi-periodic stellar oscillations
via cosine-filtering. The global astrometry step is key to this
process -- it minimizes the effect of spacecraft drift on the aperture
photometry, and allows us to accurately model the spacecraft motion
for further detrending.

In this paper, we describe our {\em K2} photometry pipeline and the high 
precision light curves from the reduction of {\em K2} Campaign 1.
We introduce our effort of deriving accurate astrometry for the {\em K2} 
observations, making use of the {\em K2} FFIs, and present a 
revised {\em K2} Campaign 1 target list in \S 2. 
In \S 3, we present our aperture photometry method.
In \S 4, we revisit our detrending techniques and present our 
\lcs\ at different detrending stages. In \S 5, we compare our 
photometry with that of other studies.   

\section{Astrometry}
\label{sec:astrometry}

\subsection{Background}

The first step of our reduction is to derive an accurate 
astrometric solution of the {\em K2} data. 
Despite its large pixel scale ($\sim 4$\arcsec) 
and PSF FWHM ($5-6$ \arcsec), the original {\em Kepler} 
mission turned out to be a great tool for accurate 
astrometry itself because of its extremely high SNR 
photometry and stable pointing.
\citet{Monet:2010} reported a preliminary astrometric solution 
precision, from the first few months of {\em Kepler} data, to 
be 0.001 pixel, nominal 4 mas. This high astrometric precision, and
high stability of the centroid position, enabled the high
photometric precision of the {\em Kepler} primary mission.

Unlike the original {\em Kepler} Mission, the {\em K2} stars 
typically drift across the CCD plane at a speed of 1-3$\%$ of a pixel 
every 30 min. Since the 30 min {\em K2} frame is composed of 270 short 
exposures of 6\,s each, the final PSF is inevitably distorted, and 
neighbouring stars tend to become blended. Therefore, it is difficult 
to determine accurate centroids from source extraction alone. 
Thus, we use an external catalogue, namely the fourth 
United States Naval Observatory (USNO) CCD Astrograph 
Catalogue, UCAC4 \citep{Zacharias:2013}, which has an astrometric 
precision of 15-100 mas, to derive good astrometric solutions 
for the {\em K2} frames. 

A good astrometric solution does not only benefit the 
photometric precision, but also enable us to make maximal 
use of the {\em K2} observations. 
The {\em K2} campaigns observe targets proposed by the community, 
and each target was then assigned a stamp of size 20-50 pixels across. 
This stamp size is much bigger than the original {\em Kepler} stamp size. 
In addition to the target, many other sources are observed in a 
typical {\em K2} stamp. We provide position information and
reduced light curves for all of the stars observed in the {\em K2} stamps. 
We anticipate an improved planet yield from this approach, due to 
the larger number of sources available, and the availability 
of the light curves of neighbouring stars, useful in blend analyses.

We first derive a general astrometric solution using the Kepler
FFIs and our custom developed astrometry software used for HATNet. 
This general solution is
then used as an initial guess for the remaining stamp observations. We
stitch all the {\em K2} stamps together into a ``Sparse FFI'' (SFFI), with
the unobserved regions masked. We fit for an astrometric solution to
the SFFI, which is assumed to be a low order polynomial distortion from 
the FFI astrometric solution.

\subsection{Astrometry Standard Catalogue}
We use the UCAC4 catalogue \citep{Zacharias:2013} as our astrometry standard 
for deriving the astrometric solution. It contains over 113 million 
objects, and is complete down to magnitude $R=16$. 

The precision of coordinates provided by UCAC4 is $\sim 15-100$ mas. 
UCAC4 catalogue also contains proper motion of $\sim 105$ million stars, 
with errors around 1 to 10 ${\rm mas}/{\rm yr}$. Both the coordinates 
and proper motions are measured on the International Celestial Reference 
System (ICRS) at a mean epoch of 2000. We linearly corrected the 
coordinates based on the proper motions to epoch 2014. UCAC4 
also contains Two Micron All-Sky Survey 
(\citep[2MASS;][]{Skrutskie:2006}) photometry for 
around 110 million stars, and AAVSO Photometric All-Sky 
Survey (APASS) five-band (BVgri) 
photometry for over 51 million stars. For the stars in 2MASS
but without APASS photometry, their $gri$ band photometry are
estimated using the 2MASS magnitudes. BV band magnitudes are adopted
from the Tycho-2 catalogue where available, otherwise also estimated from 2MASS.
We use the B and V magnitude to estimate the magnitude of stars in 
the {\em Kepler} band when needed.

\subsection{Astrometry on the Full Frame Image}
\label{sec:ffi}
The {\em Kepler} FFIs are divided into subimages by readout 
channels. There are 84 subimages for each FFI. Two of the CCD 
modules (8 channels altogether) failed during the {\em Kepler} main mission.  
The remaining 76 subimages were used to create the images from the 38
working CCDs (following the Kepler Instrument Handbook). 
We use {\sc fistar} \citep{Pal:2012} for source extraction. 
The uncertainties of the source extractor is about 0.07 pixels. This 
is estimated by comparing the extracted source positions on the two 
different FFIs taken from the Campaign 1. The relative shift and 
rotation between the two FFIs were taken into account by fitting 
a low order polynomial to the two extracted source lists.
We also experimented with other source extractors such 
as {\sc sextractor} \citep{Bertin:1996}, and all gave similar 
uncertainties.

The astrometric solution is provided by {\sc anmatch}, a 
software routinely produces arcsecond 
precision astrometric solutions for the HATNet/HATSouth observations. 
{\sc anmatch} first uses the engine of 
astrometry.net \citep{Lang:2010} for a low 
order solution (3rd order Simple Imaging Polynomial [SIP] tweak) 
to obtain an initial guess, then fits a 3rd order polynomial 
on a bigger matched list between the extracted source and the 
catalogue source to obtain the final solution. 

\begin{figure}
\includegraphics[width=\linewidth]{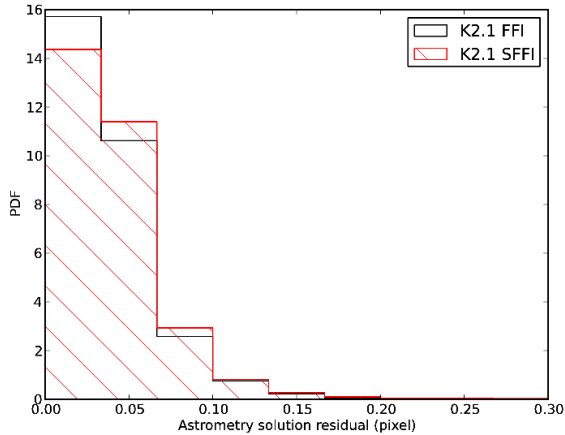}
\caption{
The residual of the astrometric solution for all channels of 
the {\em K2} Campaign 1 field FFIs (black) and SFFIs (red and hatched). 
This is computed by comparing the distance between the projected catalogue 
coordinates and the detected source coordinates on the CCDs. 
\label{fig:astrom_res}
}
\end{figure}

The histogram of the residuals from the astrometric solution, for all
38 CCDs in the {\em K2} Campaign 1 FFIs, is plotted in black in
Figure~\ref{fig:astrom_res}. The astrometric solution 
residual is defined as the distance between the projected pixel 
coordinates of catalogue sources and the corresponding coordinates 
for the same stars from our source extractor. The median of 
the astrometric solution residual for these raw 
frames is $\sim 0.032$ pixel for {\em K2} Campaign 1. \footnote{We expect 
a factor of two difference between the estimated uncertainties from the 
source extractor, and our astrometric residual, given the different methods 
by which these two uncertainties are calculated. We estimated the uncertainty 
of the source extractor by compare the .rms. difference of the source
positions on two frames (allowing a spacial transformation), while 
the uncertainty of the astrometric residual is estimated by the median of residual 
between the extracted position and the projected solution.}
Given the {\em Kepler} CCD has 
a plate scale of 3.98$^{\prime\prime}$, our astrometric solution 
residual corresponds to 0.127$^{\prime\prime}$.   

\subsection{Astrometry for All the Stamps} 
\label{sec:astrom_result}
The SFFIs are very sparse. For a single channel, typically more than 
$95\%$ of the pixels are not downloaded. It is impossible to solve 
for the astrometric solution of these SFFIs via a direct catalogue
matching. We use the FFI astrometric solution as an initial guess to 
overcome this problem. 

\begin{figure*}
\includegraphics[width=\linewidth]{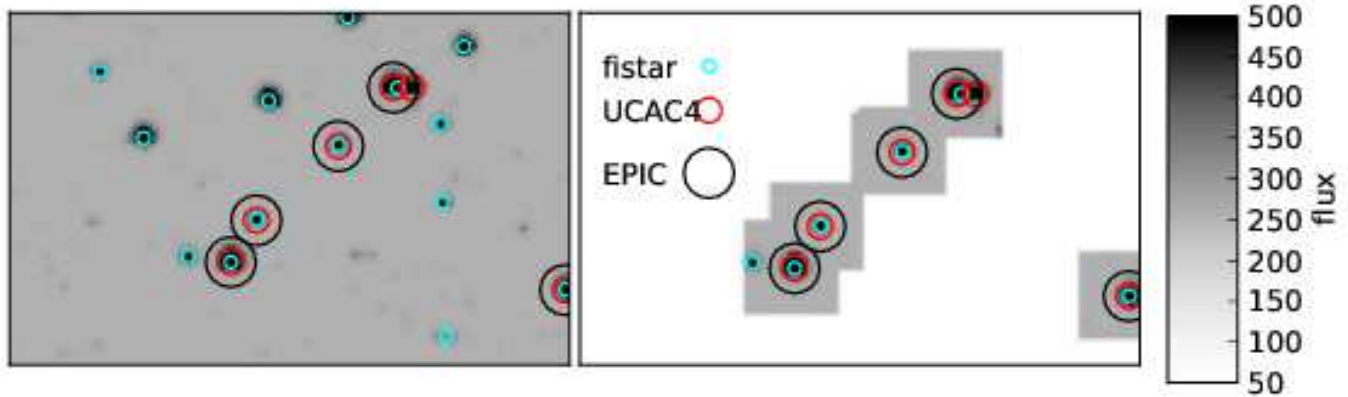}
\caption{
An 85 $\times$ 56 pixel$^2$ region of the FFI frame (left) and its corresponding 
SFFI frame (right). The detected sources are marked by cyan circles, 
and the projected sources from UCAC4 catalogue by red circles. 
The original {\em K2} targets are marked by black big circles centred 
on centroids as determined from our astrometric solution.
The white region in the SFFI image were not observed, and 
are masked out. 
\label{fig:proj}
}
\end{figure*}

We first use {\sc fistar} to extract the sources from 
the SFFIs. We then project the UCAC4 catalogue on to the SFFIs with the 
astrometric solutions obtained in \S \ref{sec:ffi}. 
We use an iterative point matching algorithm allowing a field 
centre shift from the FFI to SFFI to match the extracted sources 
and the projected coordinates of catalogue stars. 
We solve for the distortion between these matched pairs using 
a second order polynomial to obtain the final solution. 
In Figure \ref{fig:proj}, we show the 
corresponding region of FFI and SFFI from the same CCD 
channel ({\em K2} Campaign 1, module 13, channel 41). This region 
consists of three stamps in the SFFI observations. We marked out the 
detected source by cyan circles, 
and the projected sources from catalogue by red circles. 
The original {\em K2} targets are marked out in the black circles. 
Some stamps consist of multiple stars. 
We also show that in the top rightmost stamp in Figure \ref{fig:proj}, 
the projected catalogue indicates that there are additional sources 
blended in the primary source's PSF, which was originally missed 
by the source extractor, and the light of which would be 
measured together with that of the primary source. 

We compute our astrometric residuals as per \S \ref{sec:ffi}.   
The astrometric residuals on the SFFIs for {\em K2} Campaign 1 are 
shown in red in Figure \ref{fig:astrom_res}. The median astrometric 
residual is around 0.034 pixels (0.135$\arcsec$), comparable 
with what we achieved on the FFIs.

\subsection{A revised {\em K2} target catalogue} 

We projected the UCAC4 catalogue on the {\em K2} Campaign 1 SFFIs using the 
astrometric solution we obtained in \S \ref{sec:astrom_result}. Stars
with centroids within 3 pixels from the stamp edges were excluded. We also 
included those stars in the original 
Ecliptic Plane Input Catalogue (EPIC, \citet{Huber:2015}) 
but not in the UCAC4 catalogue. The EPIC catalogue is a combination 
of the Hipparcos catalogue \citep{van_Leeuwen:2007}, 
Tycho-2 catalogue \citep{Hog:2000}, UCAC4 catalogue, 2MASS, and
SDSS DR9 \citep{Ahn:2012} for the selected {\em K2} target stars. 
There may be 
systematic offsets between the coordinates from the above catalogue, 
but they are relatively small ($\sim$ 10 mas), and can be ignored 
when combining these catalogues.
We estimated the B and V magnitudes of stars not in the EPIC catalogue
as per the Kepler Instrument Handbook. 
Altogether, we found 14778 stars from the 
UCAC4 catalogue, and an additional 7939 stars from the EPIC catalogue
only, in {\em K2} Campaign 1. 
This combined set of {\em K2} target catalogue 
(22717 stars in total) is larger by 5$\%$ than 
the total in the original EPIC catalogue 
(21647 stars in total). 
This increase will be more pronounced for other, more 
crowded {\em K2} fields. Part of the final {\em K2} Campaign 1 
target list catalogue is shown in Table 
\ref{table:catalog}. We provide the centroid positions on the 
corresponding postage stamp, for each target.

\subsection{The refined motion for each module}
\label{sec:scmotion}

\begin{figure}
\includegraphics[width=\linewidth]{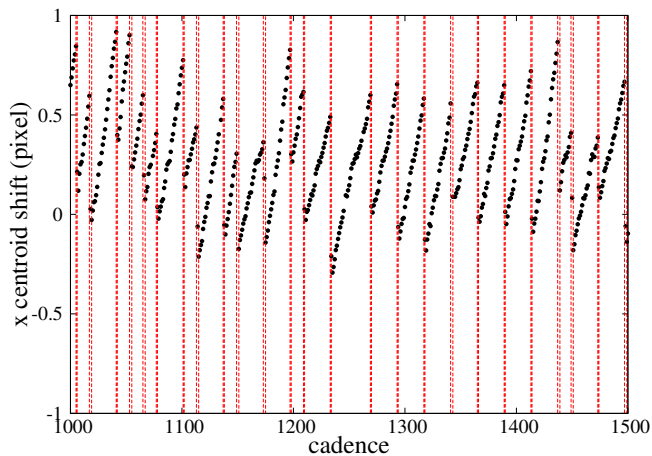}
\caption{A segment of relative $X$ centroid drift from module 4, between 
cadence 1000 and 1500. The dashed red lines separate the drifting 
segments we identified via edge detection. 
\label{fig:motion}}
\end{figure}

VA14 pointed out that the {\em K2} photometry is strongly 
correlated with the centroid positions of the stars. They also 
found that the centroid position of individual stars, as determined 
by their weighted light centres, are often not good enough. 
As such, they chose the centroid 
motion of a bright star to represent all the stars observed in the 
same campaign. Taking advantage of our derived astrometric solution, 
we find that by combining many stars observed on the same module, 
we can achieve even better constrained $X$,$Y$ motion tracks. 
We define the $X$, $Y$ motion derived for the centre pixel position 
of SFFI modules as the refined motion for each module. 
As an example, we show in Figure \ref{fig:motion} the 
relative $X$ centroid drift of module 4, a module in the corner of 
the focal plane. 
We did not derive a rolling motion for the entire spacecraft 
to avoid correcting for additional rotations between modules. 
We notice that although the spacecraft attempted to correct its 
roll drift every 12 hours, the drifting segments can 
last longer. The drifting segments are defined as a time 
series of smooth $X$, $Y$ motion without significant outliers.  
We identify each drifting segment, and the outliers in between segments,
by applying a 1-d edge detection method (Sobel operator) on the $X$ 
motion of each module. A example of the edge detection is show 
for module 4, in Figure \ref{fig:motion}, with the 
red dashed lines separating each segment. 

\section{Photometry}
\label{sec:photometry}

For each target, we use {\sc fiphot} \citep{Pal:2012} 
to extract photometry in 36 circular apertures around the derived 
centroids. 
The flux from the sources are estimated by summing up all the 
pixels within an aperture and weighting edge pixels by the 
fraction of which lie within the circular aperture. The background 
flux is measured by taking the median, with iterative outlier 
rejection, in an annulus of pixels around the aperture, then
multiplying it by the area of the aperture, and 
subtracting it from the flux. The aperture sizes range 
from 2.5 --- 5 pixels, and are chosen so as to optimize 
the photometric precision for a wide range of magnitudes. 
For apertures with sizes smaller (larger) than 4.5 pixels, the 
background annulus has inner radii of 5 (6) pixels and outer 
radii of 11 (12) pixels.  
They are designed to optimize the photometric precision for stars in 
different magnitude bins. 
We note that for saturated stars (KepMag $<10$), 
our photometric method cannot capture all the leaked 
electrons in the stamps, therefore leading to degraded 
photometry. 
For the saturated stars, the fixed aperture approach 
taken by VA14 remains the best way to extract optimized 
photometry for now.

\section{Light Curves and Detrending}
\label{sec:light curves}

We present the raw aperture photometry light curves (described in 
section \S \ref{sec:photometry}), and apply a three-step detrending 
process on our \lcs. The detrending methodology is adapted from the
HATNet pipeline, as well as the {\em Kepler} 
light curve detrending pipeline described in \citet{Huang:2013}. 
Each step of detrending is aimed to correct different aspects of the 
noise in the \lcs . Users are able to query light curves detrended 
up to an intermediate step to suit their own purpose 
\footnote{http://k2.hatsurveys.org/}. In this section, 
we will first describe the properties of our detrending methods, and then 
demonstrate the \lc\ products from each detrending step. At the end of 
this section, we will compare our \lcs\ with those from other works.

\subsection{Detrending}
We refer to our \lcs\ from the aperture photometry as \rlc.   
Our detrending pipeline applied to these \rlc\ can be divided into the 
following three steps:

(1) External Parameter Decorrelation (EPD);

(2) Trend filtering (TFA);

(3) Cosine filtering (COS). 

To correct for the photometric variations due to the motion of the
spacecraft, we performed EPD on the \rlc\ \citep{Bakos:2010}. 
We follow a similar methodology 
as described in VA14 
to deal with the thrust fire events of the spacecraft. 
Instead of correcting for the drift effect due to a 6 hour roll, 
we make use of the drifting pattern 
of each module we identified from \S \ref{sec:scmotion}. We first 
reject the data points that fall in between any drifting segments. 
We then divide the data into 6 segments before detrending, as defined in 
Table \ref{table:seg}. These segments are designed to separate large 
amplitude flux offset in the data, and allowing each segments to 
be represented by low order smooth functions. 
For each segment, we iteratively fit a 3rd order B-spline through 
the median magnitude of each drifting segment with 3-$\sigma$ outlier 
rejection until the fit converges. 
This long term trend represented by the B-spline is then removed. 

We then fit for the variation due to spacecraft drift as per the 
following:

\begin{equation}
\begin{aligned}
f(m) &= c_0+c_1\sin(2\pi\,X)+c_2\cos(2\pi\,X)\nonumber\\
     &+c_3\sin(2\pi\,Y)+c_4\cos(2\pi\,Y)\nonumber \\
     &+c_5\sin(4\pi\,X)+c_6\cos(4\pi\,X)\nonumber\\ 
     &+c_7\sin(4\pi\,Y)+c_8\cos(4\pi\,Y),
\end{aligned}
\end{equation}
in which, $X$, $Y$ represent the relative $X$, $Y$ drift of the module on
which the target sits. 
The fitted $X$, $Y$ trend is then removed from the original \rlc, and the
B-spline long term trend added back in. This preserves the long-term
trend while minimizing the effects of short-term spacecraft motion. The \lcs\ at this stage 
is called \epdlc.

The shared systematics between the stars are then corrected
using an adaptation of the Trend Filtering Algorithm (TFA) designed 
for {\em Kepler}. The idea of TFA is to select a set of template \lcs, 
that is representative of all the systematic variations present in 
the data. Each target \lc\ is then corrected based on a linear
filter that identifies the shared trends between the target and the
template \lcs. We found that using only template stars observed in 
the same channel as the target provided the best results. Since 
the number of stars observed in each channel in {\em K2} Campaign 1 is quite small, 
we use all the stars but the target as templates in 
the TFA procedure. The TFA filtered \lcs\ are 
denoted as \tfalc.

The last step is to filter all the low frequency variabilities 
(mostly due to intrinsic stellar variability) using a set of cosine 
and sine functions.  
This method was implemented by \citet{Huang:2013} for the 
independent search of planetary candidates in the 
original {\em Kepler} data. 
We aim to keep all periodicities at or below
the protected timescale of the transit undisturbed, 
while minimizing any other variations following \citet{Kipping:2013}. 
The cosine function detrending is applied to the \epdlc\ light curves, 
the resulting \lcs\ are called the \coslc. The cosine function detrending 
process is independent of the TFA process above. Due to its purpose, 
astrophysical signals such as stellar pulsations are no longer preserved 
in the \coslc.    

\subsection{Light Curve Products}

\begin{figure*}
\includegraphics[width=\linewidth]{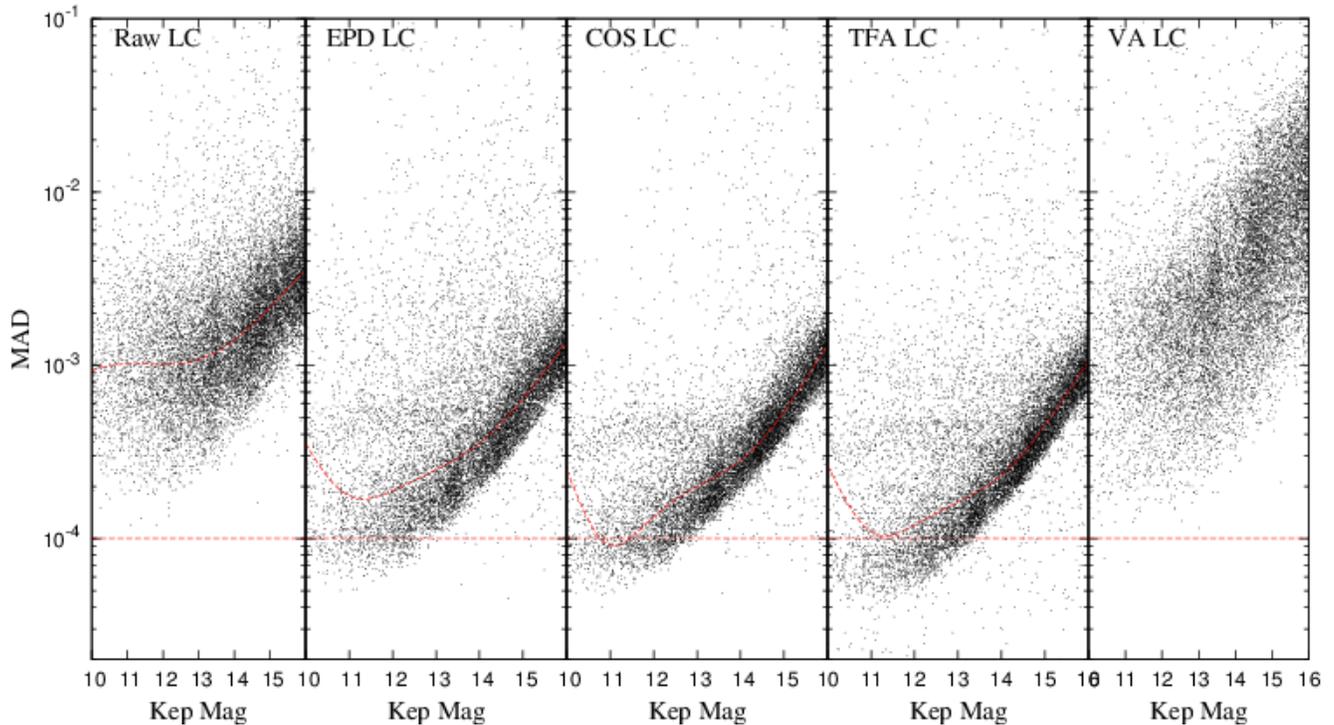}
\caption{
The point to point median standard deviation around the 
median (MAD) versus {\em Kepler} magnitude of all the light curve products 
at different detrending stages. From left to right, we show the \rlc\, 
the \epdlc, the \coslc\, the \tfalc\ and the VALC. The dashed red line 
is the fitted function for the magnitude versus the median MAD in the 
magnitude bin. The solid horizontal line indicates a scatter of $10^{-4}$ (100 ppm).
The vertical scale is logarithmic, and is the same for each panel. 
\label{fig:mmd_self}
}
\end{figure*}

\begin{figure*}
\includegraphics[width=\linewidth]{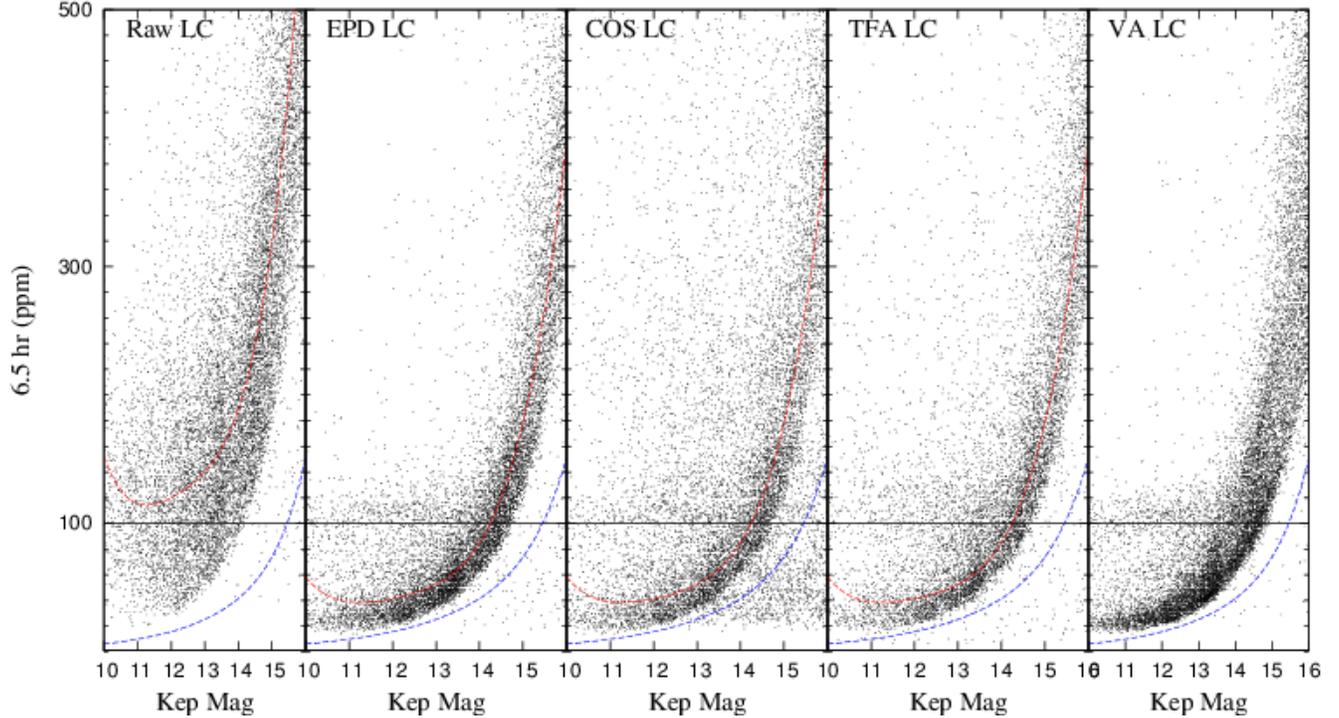}
\caption{
    The 6.5 hour precision versus {\em Kepler} magnitude of all the light curve 
products at different detrending stages. From left to right, we 
show the \rlc\, the \epdlc, the \coslc\, the \tfalc\ and the VALC. 
The solid black horizontal line indicates a scatter of $10^{-4}$ (100 ppm). 
The vertical scale are the same for each panel, linear and in units of parts 
per million (ppm). The dashed red line is the fitted function 
for the magnitude versus the median 6.5 hour precision in the magnitude bin.
The dashed blue line is the fitted function indicate the bottom 
envelope of the original {\em Kepler} 6.5 
Hour precision based on \citet{Jenkins:2010}.
\label{fig:65h_self}
}
\end{figure*}

We provide two types of measurements about the precision of our \lcs. 
We use the point to point median scatter around the median (MAD) to 
represent the overall variability in the \lcs\, which is used as an 
estimation of noise level in our transit search algorithm. We also report 
the 6.5 hour precision as per VA14, which characterizes the noise of \lcs\ 
at a time scale relevant to the transit duration of an earth analog. 
Figure \ref{fig:mmd_self} shows the MAD of our \lcs. The best 
precision of \rlc, \epdlc, \coslc\ and \tfalc\ for the bright stars are 
$1.2\times10^{-4}$ (120 ppm), 
$\sim 6\times10^{-5}$ (60 ppm), $\sim 5\times10^{-5}$ (50 ppm) 
and $\sim 5\times10^{-5}$ (50 ppm), respectively. 
Figure \ref{fig:65h_self} shows the estimated 6.5 hour 
precision of our \lcs. The 
best precision of \rlc\ for the bright stars are 
$\sim 2\times10^{-5}$ (20 ppm), and $\sim 1.5\times10^{-5}$ (15 ppm) 
for all the other three types of \lcs. We also overlaid the 
estimation of the bottom envelop of the original {\em Kepler} 6.5 Hour 
precision based on \citet{Jenkins:2010}.

\begin{figure*}
\includegraphics[width=\linewidth]{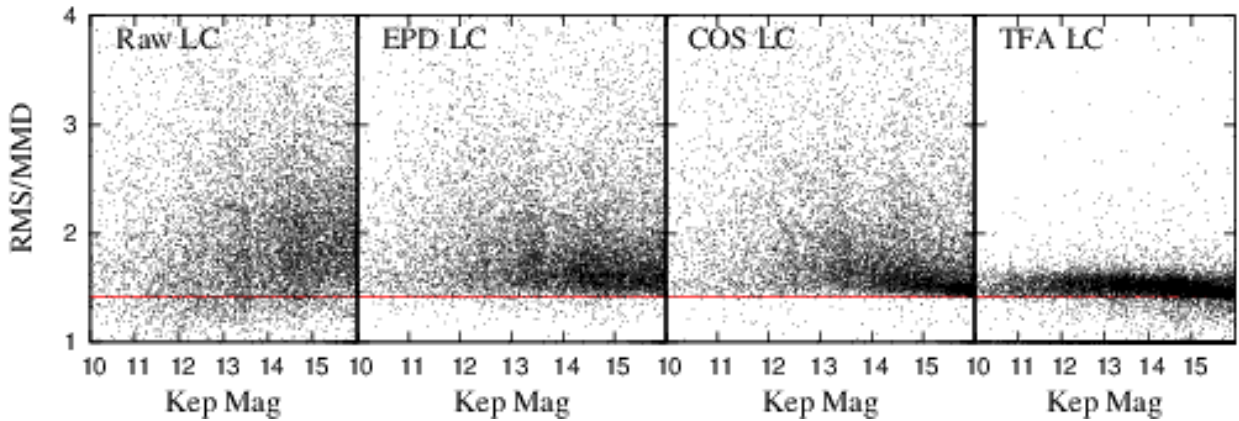}
\caption{
    Ratio of point to point RMS and MAD versus {\em Kepler} magnitude. 
From left to right, we show the \rlc\, the \epdlc, 
the \coslc\ and the \tfalc. The red
horizontal line indicates the value of $\sqrt{2}$, which 
should be the value or their ratio for pure white noise.
\label{fig:cmprms}
}
\end{figure*}

We show that the EPD process always improves 
both the short time scale (6.5 hours) and 
the long time scale (whole campaign) 
precision compared to the \rlc. We find a greater improvement for the
bright stars compared to the faint stars. The COS filtering 
process improves the precision in both time scales compared to \epdlc. 
For most stars, the \tfalc\ has a similar or marginally 
worse precision compared with the \coslc, but the TFA 
process tends to preserve the intrinsic 
variabilities of the stars. We note that for a small fraction of 
the stars, the \coslc\ and \tfalc\ can have a worse 6.5 hour precision 
compared to the \epdlc. This is because the cosine filter and TFA 
algorithm both use linear least square method aiming to minimize the 
overall point-to-point scattering in the \lcs, which is sometime 
achieved at the cost of increasing noise at specific time scale. 
We compare the noise properties of 
the \lcs\ in Figure \ref{fig:cmprms} by showing the ratio of 
per point root mean square (RMS) and MAD versus magnitude. 
If the noise is composited with pure white noise, this ratio should 
be $\sqrt{2}$. The \tfalc\ have the most white noise compare to 
other detrending stages.

\begin{figure*}
\includegraphics[width=\linewidth]{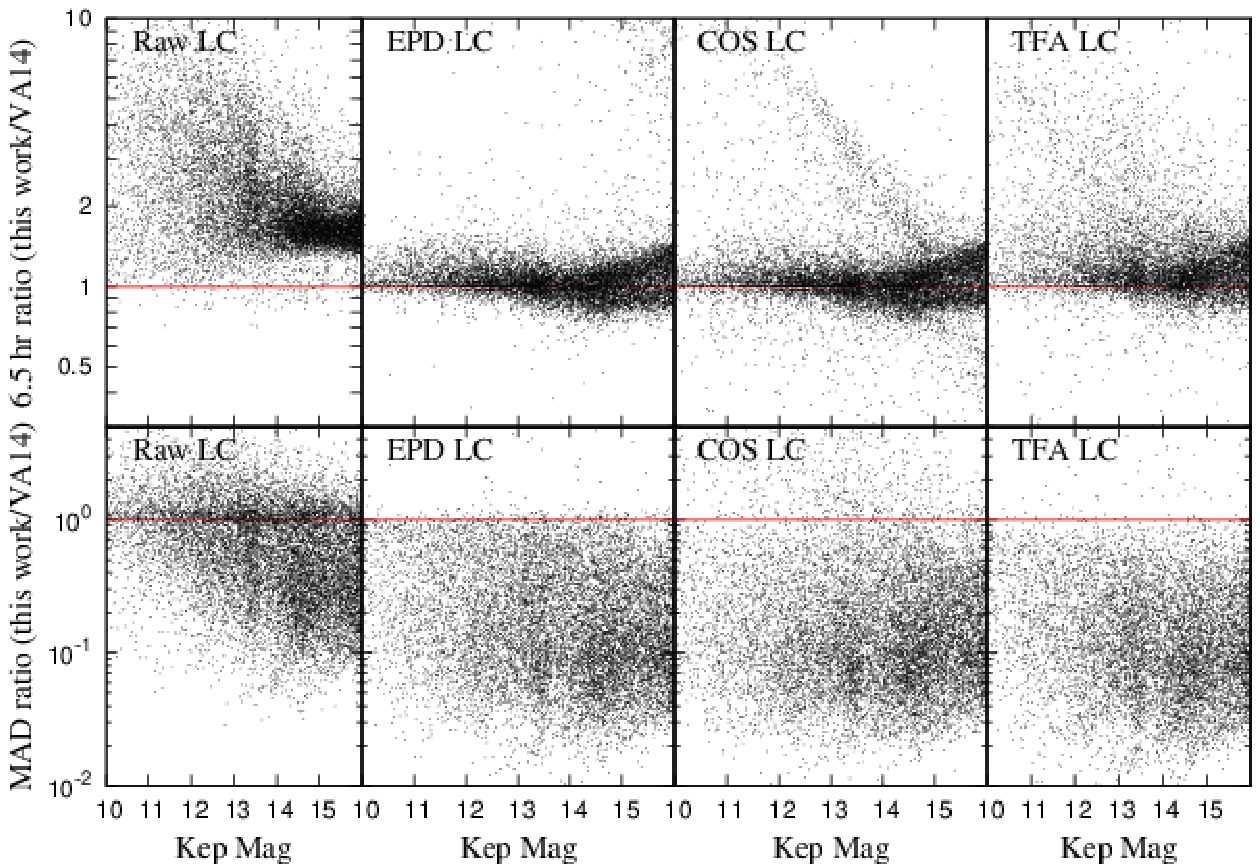}
\caption{
Photometry precision of our \lcs\ compared to VA14 {\em K2} 
Campaign 1 \lcs. Top panel: 6.5 hour precision ratio 
between our \lcs\ and VA14 \lcs\ versus {\em Kepler} magnitude. 
Bottom panel: Per point MAD ratio between our \lcs\ and 
VA14 \lcs\ versus {\em Kepler} magnitude. 
From left to right, we show the \rlc\, the \epdlc, 
the \coslc\ and the \tfalc. The red horizontal line 
indicates the value of 1. We note the vertical scale in 
the two panels are different, and both in log scale.
\label{fig:cmpwithVA}
}
\end{figure*}

\section{Comparison with other works}

Many teams have developed methods to improve {\em K2} photometry. 
We summarize the different approaches according to their photometry and 
detrending methods. 

\subsection{Photometry Methods of previous works}

All the teams use aperture photometry method to extract the light curves 
from {\em K2} data. However, they differ in the details of aperture choice 
and centroid measurements:

\begin{itemize}

\item Fixed Mask method:

The fixed mask method is such that the flux of the target 
is summed up over pixels within a fixed predetermined mask, while the 
pixels are accounted in a binary way. 
VA14 %\citet{VanderburgJohnson:2014} 
used a combination of approximate circular 
aperture and fitted apertures using the {\em Kepler} Pixel Responding 
Function \citep{Bryson:2010}. 
\citet{Foremanmackey:2015} and \citet{Angus:2015} used approximate 
circular but binary apertures (do not include pixels partially) 
and present the photometry from the 
best apertures. \citet{Lund:2015} used the density-based spatial 
clustering of applications with noise routine 
(DBSCAN, \citet{Ester:1996}) to chose 
their pixel mask (aperture).

\item Moving Circular Aperture method: 

\citet{Aigrain:2015} used 6 circular apertures to extract photometry. 
The apertures are soft-edged, in the sense that pixels straddling the 
edge of the aperture contribute partially to the flux.

\item Centroids from astrometric solution: 

\citet{Aigrain:2015} used centroids derived from their own astrometric 
solution with the 2MASS all-sky point-source catalogue.

\item Centroids from WCS header: 

\citet{Foremanmackey:2015} and \citet{Angus:2015} used centroids from the WCS 
header of the {\em K2} target pixel files. Only one WCS solution is 
given for the entire time series of each star.

\item Centroids from weighted centre of flux:

VA14 %\citet{VanderburgJohnson:2014} 
and \citet{Lund:2015} used the 
weighted centre of flux as the centroids of the stars. We note, in 
the subsequent detrending, 
VA14 %\citet{VanderburgJohnson:2014} 
used the centroid of star EPIC 201611708 
instead of the centroids of individual stars.

\end{itemize}

In this work we used 36 moving circular apertures, with the 
centroids of apertures determined by precise astrometric solutions, to 
determine the photometry of each star.

\subsection{Detrending Methods of previous work}

There are three different types of ``detrending" methods used 
by other authors. 

\begin{itemize}
\item Decorrelation:

VA14 %\citet{VanderburgJohnson:2014}
, \citet{Armstrong:2014} and \citet{Lund:2015} used a
self-flat-field method to decorrelate the aperture photometry 
from centroid position of the image. There are, however, 
subtle differences between these studies. 
VA14 %\citet{VanderburgJohnson:2014} 
used the centroids from a representative star, 
\citet{Armstrong:2014} seemed to use centroids for individual 
stars, while \citet{Lund:2015} used 
the weighted light centroids derived for individual stars. 
VA14 %\citet{VanderburgJohnson:2014} 
used a 1-d decorrelation along 
the trajectory of the drift, \citet{Armstrong:2014} used 2-d 
centroid surface to decorrelate with the flux, and \citet{Lund:2015} 
used both the 1-d and 2-d approach in their pipeline. 

\item Gaussian Process:

\citet{Aigrain:2015} and \citet{Crossfield:2015} used Gaussian 
process model, with the rolling angle as the input variable to 
detrend the \lcs. They assume the systematics can be modelled as a 
function form of the rolling angle, and that function's form can vary 
from star to star.

\item Not Detrending:

\citet{Foremanmackey:2015} and \citet{Angus:2015} choose to not 
detrend their 
light curve prior to the search of signal, but instead, they 
simultaneously fit for the systematics and the signal of interest. 

\end{itemize}

In this work, we applied three stages of detrending. In the first 
stage, we applied a similar method as the decorrelation detrend 
in VA14. We additionally applied TFA and COS filtering to 
further filter the data. TFA is aimed to correct for 
shared systematics between the stars observed on the same 
channel, while preserving the stellar variability. The COS 
filtering method aimed to correct for any variability in 
the \lcs\, and is optimized for searching for transit signals. 

\subsection{Centroids Determination}
We took a similar approach as \citet{Aigrain:2015} in the 
determination of centroids by deriving an astrometric solution 
for each image. We made use of a more precise catalogue, UCAC4 
instead of 2MASS, and the Full Frame Image as a better 
initial guess, and achieved higher precision in our astrometric
solution. \citet{Aigrain:2015} reported a typical root mean 
square of the astrometric solution of 0.4\arcsec, or 
approximately 0.1 pixel, $\sim 3$ times larger than our typical 
astrometric residuals.  

\subsection{Photometric Precision}

To date, only VA14 %\citet{VanderburgJohnson:2014} 
have released their detrended {\em K2} Campaign 1 \lcs, therefore we 
will focus on comparing our photometry precision with their work.

We show in Figure \ref{fig:cmpwithVA} the precision ratio between 
our \lcs\ and the VA14 \lcs, for the same stars at both times 
scales. The \epdlc, \coslc\ and \tfalc\ from this work have 
comparable precision compare to VA14 \lcs\ on the 6.5 
hour time scale, and a smaller point-to-point 
scatter over the entire observation length. 

\citet{Aigrain:2015} presented their $\sigma_{\rm MAD}$ 
(similar to MAD) and 6.5 hour Combined Differential Photometric 
Precision (CDPP) for {\em K2} engineering data 
photometry. Their best precision for $\sigma_{\rm MAD}$ is $\sim$ 
300 ppm for the bright stars, and 60 ppm for 6.5 hour CDPP. 
In this work, we achieved higher precision on both time 
scales (50 ppm and 15 ppm, respectively). Although, we 
caution that the noise characteristic for {\em K2} 
engineering data and {\em K2} Campaign 1 data could be 
different. 

\subsection{Power Spectrum}

Previous works \citep{Lund:2015,Angus:2015} noted the detrended \lcs\ 
from a 1D decorrelation may still have residual spikes 
around the harmonics of $\sim 47.2271$\ $\mu\mathrm Hz$ in their power spectra. 
These residuals may be largely due to aliasing of the low 
frequency power, induced by data gaps from rejected points during thruster 
firing. These harmonics are less obvious in our light curves.

\begin{figure}
\includegraphics[width=\linewidth]{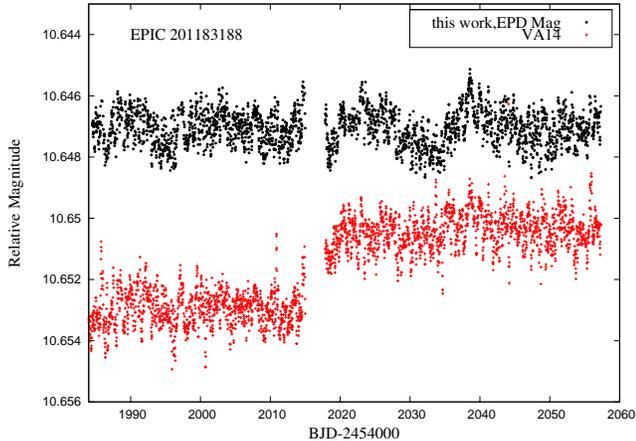}
\caption{
The \epdlc\ for EPIC 201183188 from this work (black), and 
VA14 (red). The DFT power spectrum of this star is shown in 
Figure \ref{fig:dft_dft}. 
\label{fig:dft_lc}
}
\end{figure}

\begin{figure}
\includegraphics[width=\linewidth]{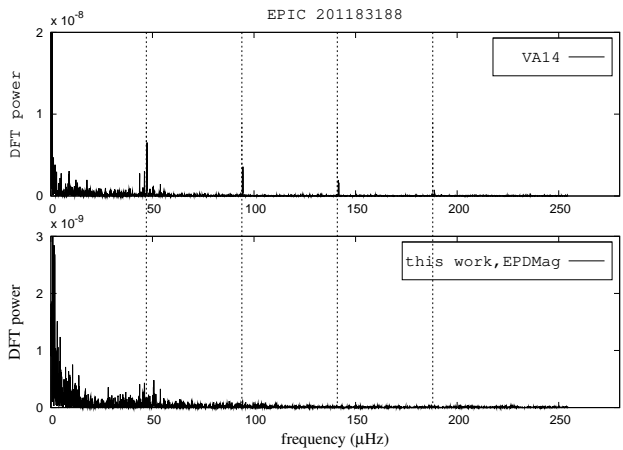}
\caption{
The DFT power spectrum for EPIC 201183188 from this work (top panel) 
and VA14 (bottom panel). Note the vertical scales 
are different in the two panels. The dash lines indicate 
the corresponding frequency associated with the spacecraft 
roll motion. For more discussion, see \citet{Angus:2015}.
\label{fig:dft_dft}
}
\end{figure}

We computed the Discrete Fourier Transformation (DFT) power spectrum using {\sc vartools} 
\citep{Hartman:2008} for the star EPIC 201183188, following
\citet{Angus:2015}. We find our \epdlc\ 
does not show the frequencies corresponding to the 6 hour roll in 
the DFT power spectrum present in the
VA14 light curves. Figure \ref{fig:dft_lc} compares our \epdlc\ 
for EPIC 201183188 with the VA14 reduction, and Figure \ref{fig:dft_dft} 
compares the DFT power spectrum of our \epdlc\ with that of
the VA14 light curve. The offset around epoch BJD 2456015 in the VALC 
could be blamed for contributing to noise peaks corresponding to the space craft 
rolling frequency.

\begin{figure}
\includegraphics[width=\linewidth]{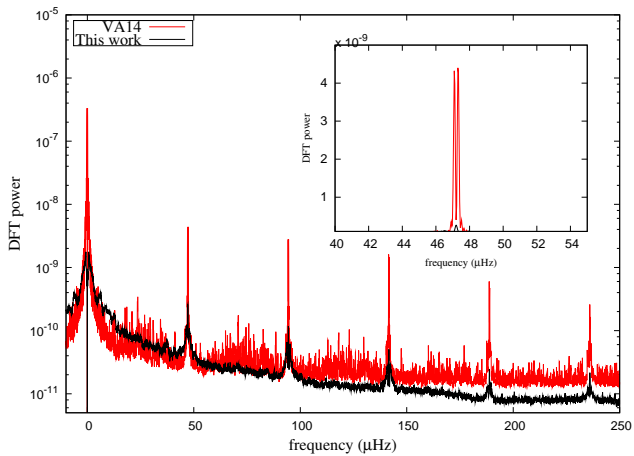}
\caption{
The median DFT power spectrum of 1661 stars in the magnitude 
range 10-12. Red dashed lines show the result from the VA14 \lcs. 
Black solid lines show the result from the \epdlc\ of this work. 
In the small window of this figure, we show the zoom-in of these 
spectra around the space craft roll frequency 
$\sim 47.2271\mu\ {\rm Hz}$. 
\label{fig:dft_all}
}
\end{figure}

We further investigate the noise characteristic of the \lcs\ by computing 
the median of DFT power spectra of 1661 stars in the magnitude 
range of 10--12. To eliminate the influence from the long term trend, we 
first filter out the strongest low frequency peak in the \lcs, and 
then recompute the DFT power spectra of each star before taking the 
median. We show this median spectrum in Figure \ref{fig:dft_all}. 
The peaks related to the rolling frequency are still present, but 
have less power compared to the median DFT spectrum computed with 
the same set of stars from VA14. When the signal from stellar 
variability is strong, such as in the case of EPIC 201183188, the 
systematic noise peaks are negligible. 

\subsection{Example Light Curves}

To demonstrate the products of our photometric extraction 
and detrending in the context of stellar variability and
transit searches, we show example light curves for stars with known variability 
and transiting signals.

\begin{figure*}
\centering
\includegraphics[width=0.8\linewidth]{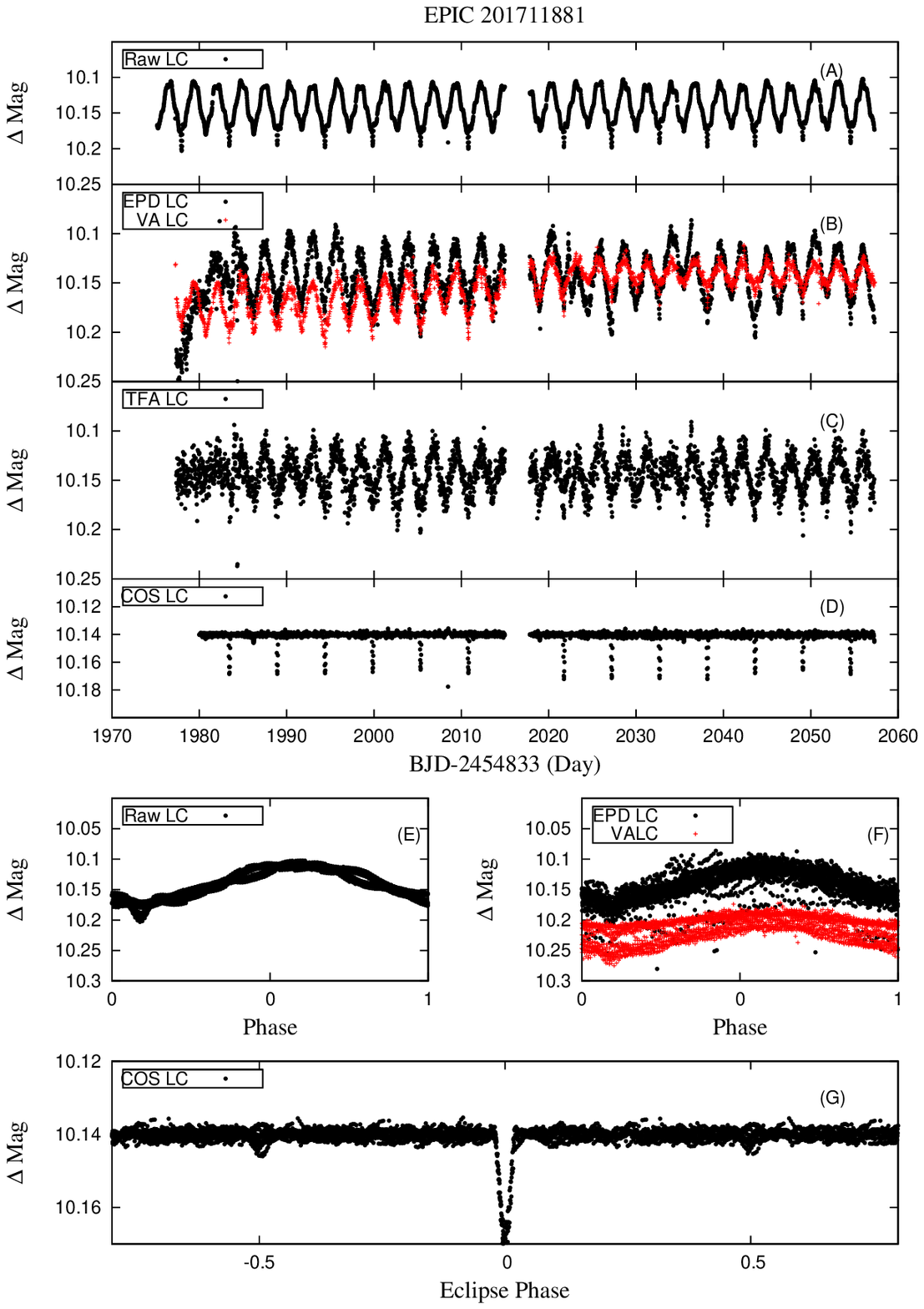}
\caption{
The \lcs\ of EPIC 201711881, a Cepheid Candidate 
discovered by \citet{Schmidt:2009}. From the top to bottom, 
we show the \rlc\ (A), \epdlc\ (B), \tfalc\ (C) and \coslc\ (D). 
In panel (E) and (F), we show \rlc\ and \epdlc\ folded 
with the period of the pulsation period ($\sim$2.735 day) in 
the phase space. The last panel we show \coslc\ (G) folded 
with twice the period of the pulsation. The eclipse 
events can be seen at phase 0. The \coslc\ in panel 
(D) and (E) have been reconstructed after the discovery of the 
eclipsing signal. Only the out-of-eclipse part have been filtered 
by the cosine filters in order to preserve the signal. 
The red \lc\ in figure (B) and (F) 
is from VA14. We note that the vertical scales are different for each panel.
\label{fig:ASAS152}
}
\end{figure*}

{\bf EPIC 201711881}, is a Cepheid candidate discovered by the 
ASAS project \citep{Schmidt:2009}. The original discovery paper 
reported a period of 2.7353$\pm5\times10^{-4}$ days. The period we 
detected is $\sim2.735$ day, consistent with \citep{Schmidt:2009}. 
The \rlc\ and \epdlc\, folded with the Cepheid 
period, are shown in Figure \ref{fig:ASAS152}. Our \rlc\, 
without any detrending, already shows a clean periodic 
signal. In addition,
our \epdlc\ preserves the amplitude of the pulsation after removing the 
systematics. We also find a periodic eclipsing signal in 
the \lc\ with twice the pulsation period. This eclipsing 
signal can be visually seen, even in the \rlc. The \coslc, phase folded
with the eclipse period, is also plotted in the bottom panel of Figure 
\ref{fig:ASAS152}. It seems that this system is more likely to be an 
eclipsing binary system with large out-of-transit variation rather 
than a real Cepheid. To further characterize this signal, the \coslc\ 
from the general detrending pipeline is not good enough. 
The \coslc\ we show in Figure \ref{fig:ASAS152} have been 
reconstructed after the discovery of the 
eclipsing signal. Only the out-of-eclipse part have been filtered 
by the cosine filters in order to preserve the shape and amplitude of 
the eclipsing signal. 

\begin{figure*}
\includegraphics[width=\linewidth]{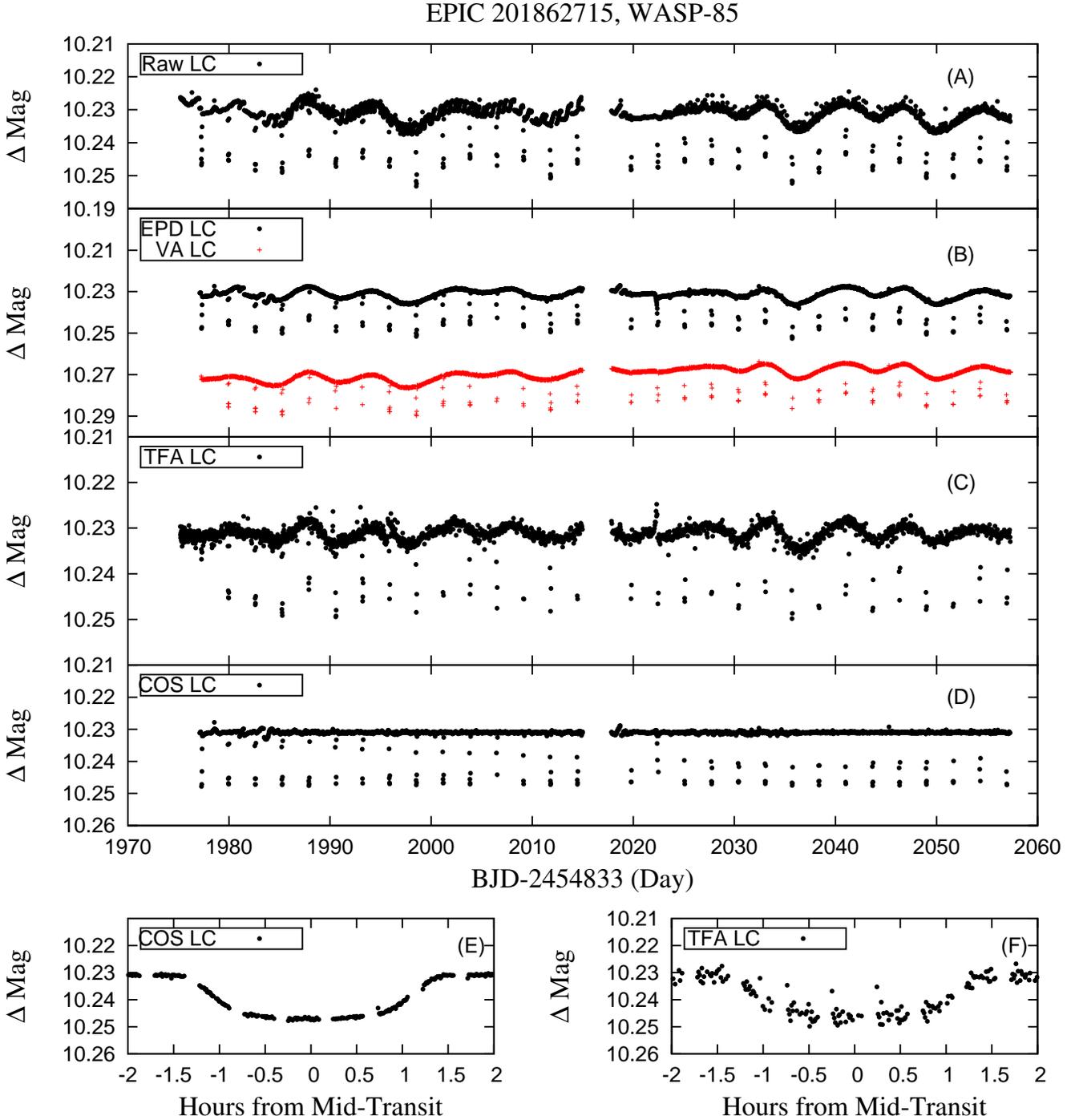}
\caption{
The \lcs\ of WASP85 (black). From the top to bottom, we show the \rlc, 
\epdlc, \tfalc\ and \coslc. The bottom most panel, we show 
\coslc\ and \tfalc\ of WASP85 folded with the epoch and period of 
WASP-85b \citep{Brown:2014}. The red \lc\ in the 
second panel is from VA14. We note that the vertical scales 
in different panels are different. 
\label{fig:wasp-85b}
}
\end{figure*}

{\bf WASP-85b}: WASP-85 (EPIC 201852715), was observed 
in {\em K2} Campaign 1, in module 15, channel 49. 1182 other stars were 
observed in the same module. We show in Figure \ref{fig:wasp-85b} 
the \lcs\ of WASP-85 at different detrending stages from this work. 
In the second panel from the top, we also overlaid the VA14 detrended \lc. 
WASP-85b has a period of $\sim$2.65 days, and known depth of 
$\sim 1.6\%$. 
We show the WASP-85 \lc\ folded with the detected period and epoch 
in phase space for both the \coslc\ and \tfalc\ in the bottom 
of Figure \ref{fig:wasp-85b}.

\begin{figure*}
\includegraphics[width=\linewidth]{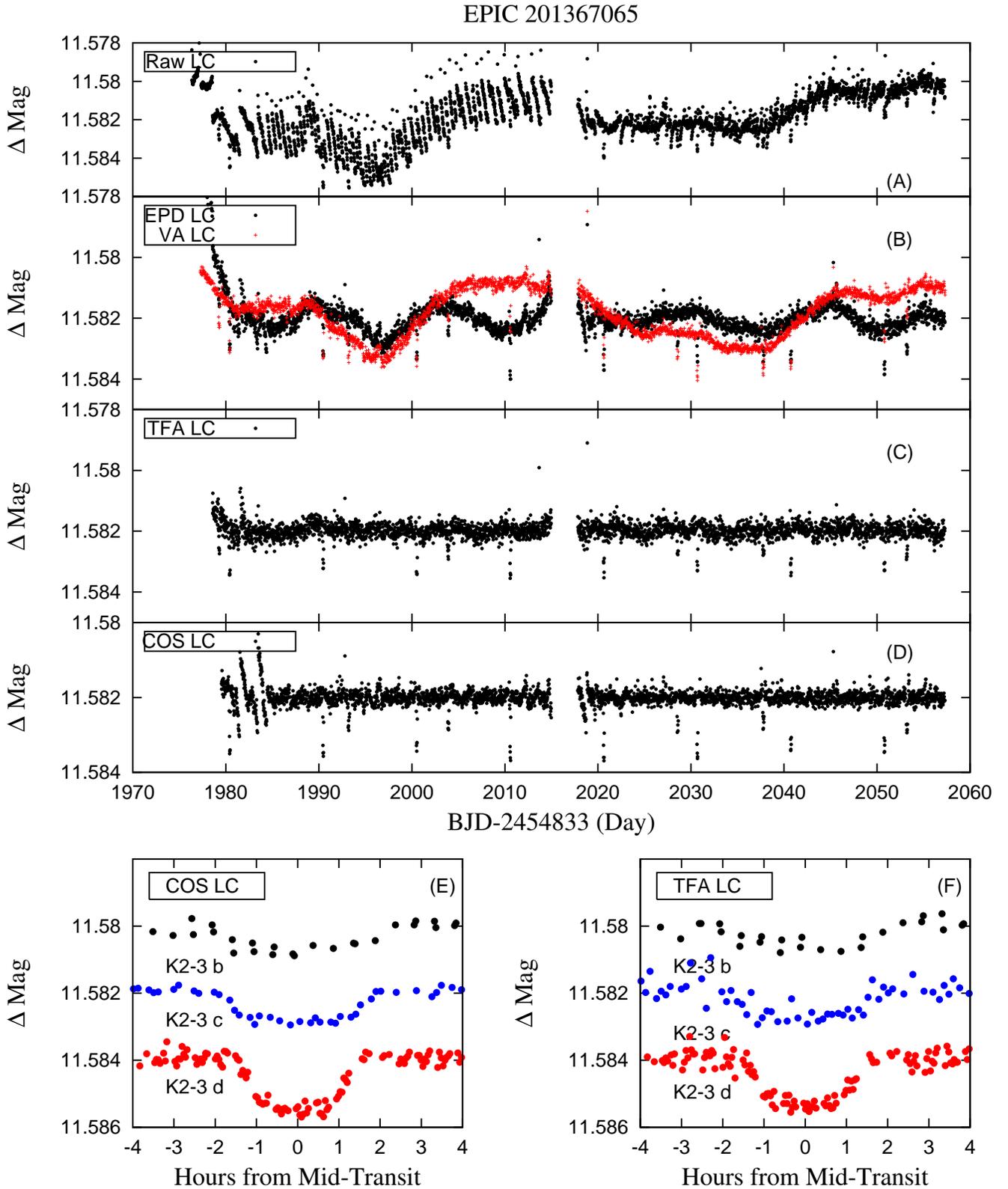}
\caption{
Same as Figure \ref{fig:wasp-85b}, but for K2-3 \citep{Crossfield:2015}. 
In the bottom panel, we show all the three planets detected in this 
system folded with their own periods and epochs. 
K2-3 b (P$\sim$ 10 day, R$\sim$2\rearth\ ), 
c (P$\sim$ 25 day, R$\sim$1.7\rearth\ ), 
d (P$\sim$ 44 day, R$\sim$1.5 \rearth\ ) are presented in red,blue 
and black curves, respectively. We note that the vertical 
scales in different panels are different.
\label{fig:K2-3}
}
\end{figure*}

{\bf K2-3}: K2-3 (EPIC 201367065), 
was observed in {\em K2} Campaign 1, in module 12, channel 40. The host star 
is an M dwarf, with three transiting super-earths discovered 
by \citet{Crossfield:2015}. We show the light curves for K2-3 in Figure 
\ref{fig:K2-3}. The transits of the biggest planet (1 mmag) is visible 
in our \rlc, and the transits of all three planets are visible in all the 
other \lcs. We also show the phase folded \coslc\ and \tfalc\ for all three 
planets in the bottom panel of Figure \ref{fig:K2-3}.

\begin{figure*}
\includegraphics[width=\linewidth]{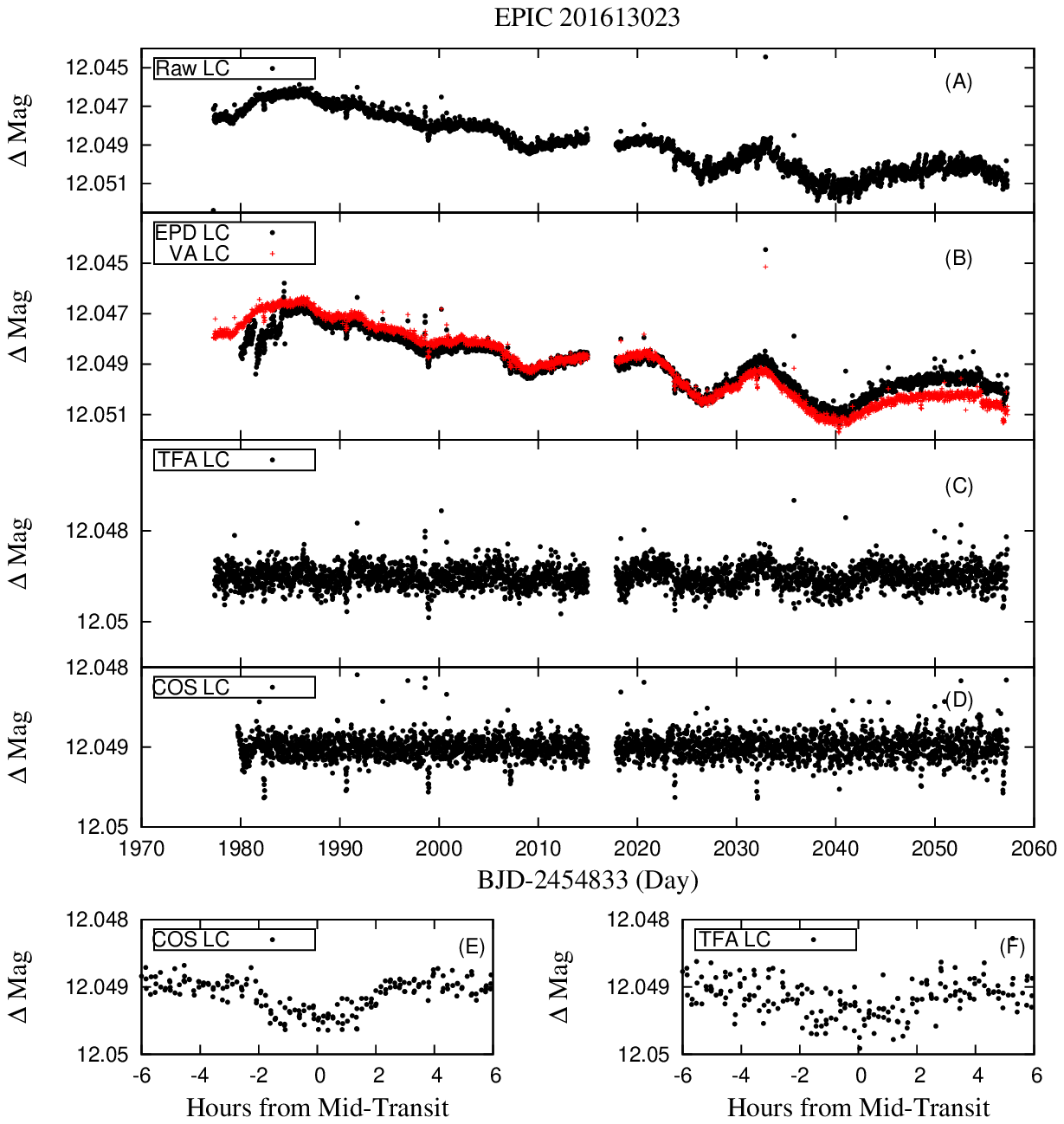}
\caption{
Same as Figure \ref{fig:wasp-85b}, but for EPIC\,201613023 
\citep{Foremanmackey:2015}. In the bottom panel we show the 
detected planet candidate folded with its period 
($\sim$8.28 days). The depth of the transit is $\sim$ 400 ppm. 
We note that the vertical scales in different panels 
are different.
\label{fig:epic201613023}
}
\end{figure*}

{\bf EPIC\,201613023}: This star was identified as a transiting
planet candidate system by \citet{Foremanmackey:2015}. We show our \lcs\ in 
Figure \ref{fig:epic201613023}. The transit signal has depth of 400 ppm. 
Individual transits from the planets are visible in the \epdlc, \tfalc\ 
and \coslc.  The phase folded \coslc\ and \tfalc\ with the detected 
epoch and period are shown in the bottom panel. 

%\clearpage

%\clearpage

%\clearpage

\section{Conclusion and Discussion}
In this article, we present our effort to extract high precision 
photometry from {\em K2} Campaign 1 data. 
Our method has three distinct 
advantages: 
\begin{itemize}

\item Making use of accurate astrometric 
solution (0.127$^{\prime\prime}$ or 0.034 pixels) from the 
FFIs for aperture centroiding; 
\item Providing photometry for all
sources on the {\em} stamps,
not only for the proposed targets from the input catalogue;
\item Presenting light curves with very low systematic variations. 

\end{itemize}

Our extracted \lcs\ are of high precision at both the long (entire 
campaign) and short (6.5 hours) time scales, even for the raw \lcs\ 
without any detrending. Light curves derived from all 36 photometric
apertures at all four detrending stages are provided for the public 
at \url{http://k2.hatsurveys.org}. 
%\clearpage

\section*{Acknowledgments}
We would like to thank the referee for their helpful comments. 
We also thank A.V for his thoughtful suggestions. 
G.Á.B. and X.H acknowledge funding from the Packard 
Foundation. This work was also supported by NASA grant 
NNX13AJ15G. K.P. acknowledges support from NASA grant NNX13AQ62G.
The {\em K2} data presented in this
paper were obtained from the Mikulski Archive for Space
Telescopes (MAST). STScI is operated by the Association of
Universities for Research in Astronomy, Inc., under NASA
contract NAS5-26555. Support for MAST for non-HST data is
provided by the NASA Office of Space Science via grant NNX09AF08G
and by other grants and contracts. This paper includes data
collected by the {\em Kepler} telescope.
Funding for the {\em K2 Mission} is provided by the NASA Science 
Mission directorate.

\bibliographystyle{apj_eprint}
%\bibliography{ref}

\begin{thebibliography}{}

\bibitem[\protect\citeauthoryear{{Ahn} et~al.}{{Ahn} et~al.}{2012}]{Ahn:2012}
{Ahn}, C.~P., {Alexandroff}, R., {Allende Prieto}, C., et~al. 2012, \apjs, 203,
  21

\bibitem[\protect\citeauthoryear{{Aigrain} et~al.}{{Aigrain}
  et~al.}{2015}]{Aigrain:2015}
{Aigrain}, S., {Hodgkin}, S.~T., {Irwin}, M.~J., {Lewis}, J.~R.,  \& {Roberts},
  S.~J. 2015, \mnras, 447, 2880

\bibitem[\protect\citeauthoryear{{Angus}, {Foreman-Mackey}, \&
  {Johnson}}{{Angus} et~al.}{2015}]{Angus:2015}
{Angus}, R., {Foreman-Mackey}, D.,  \& {Johnson}, J.~A. 2015, ArXiv e-prints,
  1505.07105

\bibitem[\protect\citeauthoryear{{Armstrong} et~al.}{{Armstrong}
  et~al.}{2014}]{Armstrong:2014}
{Armstrong}, D.~J., {Osborn}, H.~P., {Brown}, D.~J.~A., et~al. 2014, ArXiv
  e-prints, 1411.6830

\bibitem[\protect\citeauthoryear{{Bakos} et~al.}{{Bakos}
  et~al.}{2010}]{Bakos:2010}
{Bakos}, G.~{\'A}., {Torres}, G., {P{\'a}l}, A., et~al. 2010, \apj, 710, 1724

\bibitem[\protect\citeauthoryear{{Bertin} \& {Arnouts}}{{Bertin} \&
  {Arnouts}}{1996}]{Bertin:1996}
{Bertin}, E.,  \& {Arnouts}, S. 1996, \aaps, 117, 393

\bibitem[\protect\citeauthoryear{{Brown} et~al.}{{Brown}
  et~al.}{2014}]{Brown:2014}
{Brown}, D.~J.~A., {Anderson}, D.~R., {Armstrong}, D.~J., et~al. 2014, ArXiv
  e-prints, 1412.7761

\bibitem[\protect\citeauthoryear{{Bryson} et~al.}{{Bryson}
  et~al.}{2010}]{Bryson:2010}
{Bryson}, S.~T., {Tenenbaum}, P., {Jenkins}, J.~M., et~al. 2010, \apjl, 713,
  L97

\bibitem[\protect\citeauthoryear{{Crossfield} et~al.}{{Crossfield}
  et~al.}{2015}]{Crossfield:2015}
{Crossfield}, I.~J.~M., {Petigura}, E., {Schlieder}, J.~E., et~al. 2015, \apj,
  804, 10

\bibitem[\protect\citeauthoryear{{Ester} et~al.}{{Ester}
  et~al.}{1996}]{Ester:1996}
{Ester}, M., {Kriegel}, H.-p., {Sander}, J.,  \& {Xu}, X. 1996, AAAI Press, 226

\bibitem[\protect\citeauthoryear{{Foreman-Mackey} et~al.}{{Foreman-Mackey}
  et~al.}{2015}]{Foremanmackey:2015}
{Foreman-Mackey}, D., {Montet}, B.~T., {Hogg}, D.~W., et~al. 2015, \apj, 806,
  215

\bibitem[\protect\citeauthoryear{{Hartman} et~al.}{{Hartman}
  et~al.}{2008}]{Hartman:2008}
{Hartman}, J.~D., {Gaudi}, B.~S., {Holman}, M.~J., et~al. 2008, \apj, 675, 1254

\bibitem[\protect\citeauthoryear{{H{\o}g} et~al.}{{H{\o}g}
  et~al.}{2000}]{Hog:2000}
{H{\o}g}, E., {Fabricius}, C., {Makarov}, V.~V., et~al. 2000, \aap, 355, L27

\bibitem[\protect\citeauthoryear{{Howell} et~al.}{{Howell}
  et~al.}{2014}]{Howell:2014}
{Howell}, S.~B., {Sobeck}, C., {Haas}, M., et~al. 2014, \pasp, 126, 398

\bibitem[\protect\citeauthoryear{{Huang}, {Bakos}, \& {Hartman}}{{Huang}
  et~al.}{2013}]{Huang:2013}
{Huang}, X., {Bakos}, G.~{\'A}.,  \& {Hartman}, J.~D. 2013, \mnras, 429, 2001

\bibitem[\protect\citeauthoryear{{Huber} \& {Bryson}}{{Huber} \&
  {Bryson}}{2015}]{Huber:2015}
{Huber}, D.,  \& {Bryson}, S.~T. 2015, KSCI-19082-008

\bibitem[\protect\citeauthoryear{{Jenkins} et~al.}{{Jenkins}
  et~al.}{2010}]{Jenkins:2010}
{Jenkins}, J.~M., {Caldwell}, D.~A., {Chandrasekaran}, H., et~al. 2010, \apjl,
  713, L120

\bibitem[\protect\citeauthoryear{{Kipping} et~al.}{{Kipping}
  et~al.}{2013}]{Kipping:2013}
{Kipping}, D.~M., {Hartman}, J., {Buchhave}, L.~A., et~al. 2013, \apj, 770, 101

\bibitem[\protect\citeauthoryear{{Lang} et~al.}{{Lang}
  et~al.}{2010}]{Lang:2010}
{Lang}, D., {Hogg}, D.~W., {Mierle}, K., {Blanton}, M.,  \& {Roweis}, S. 2010,
  \AJ, 137, 1782, arXiv:0910.2233

\bibitem[\protect\citeauthoryear{{Lund} et~al.}{{Lund}
  et~al.}{2015}]{Lund:2015}
{Lund}, M.~N., {Handberg}, R., {Davies}, G.~R., {Chaplin}, W.~J.,  \& {Jones},
  C.~D. 2015, \apj, 806, 30

\bibitem[\protect\citeauthoryear{{Monet} et~al.}{{Monet}
  et~al.}{2010}]{Monet:2010}
{Monet}, D.~G., {Jenkins}, J.~M., {Dunham}, E.~W., et~al. 2010, ArXiv e-prints,
  1001.0305

\bibitem[\protect\citeauthoryear{{P{\'a}l}}{{P{\'a}l}}{2012}]{Pal:2012}
{P{\'a}l}, A. 2012, \mnras, 421, 1825

\bibitem[\protect\citeauthoryear{{Schmidt} et~al.}{{Schmidt}
  et~al.}{2009}]{Schmidt:2009}
{Schmidt}, E.~G., {Hemen}, B., {Rogalla}, D.,  \& {Thacker-Lynn}, L. 2009, \aj,
  137, 4598

\bibitem[\protect\citeauthoryear{{Skrutskie} et~al.}{{Skrutskie}
  et~al.}{2006}]{Skrutskie:2006}
{Skrutskie}, M.~F., {Cutri}, R.~M., {Stiening}, R., et~al. 2006, \aj, 131, 1163

\bibitem[\protect\citeauthoryear{{van Leeuwen}}{{van
  Leeuwen}}{2007}]{van_Leeuwen:2007}
{van Leeuwen}, F., ed. 2007, Astrophysics and Space Science Library, Vol. 350,
  {Hipparcos, the New Reduction of the Raw Data}

\bibitem[\protect\citeauthoryear{{Vanderburg} \& {Johnson}}{{Vanderburg} \&
  {Johnson}}{2014}]{VanderburgJohnson:2014}
{Vanderburg}, A.,  \& {Johnson}, J.~A. 2014, \pasp, 126, 948

\bibitem[\protect\citeauthoryear{{Zacharias} et~al.}{{Zacharias}
  et~al.}{2013}]{Zacharias:2013}
{Zacharias}, N., {Finch}, C.~T., {Girard}, T.~M., et~al. 2013, \aj, 145, 44

\end{thebibliography}

\clearpage

%\begin{landscape}
\begin{turnpage}
\begin{deluxetable*}{cccccccccccccccc}
\tablewidth{0pc}
\tablecaption{K2 target list\label{table:catalog}}
\tablehead{
\colhead{UCAC4ID} & \colhead{RA} & \colhead{DEC} & \colhead{J} & \colhead{H} & \colhead{K} & \colhead{B} & \colhead{V} & \colhead{g} & \colhead{r} & \colhead{i} &  \colhead{x($t_0$) \tablenotemark{b}} & \colhead{y($t_0$) \tablenotemark{b}} & \colhead{channel \tablenotemark{c}} & \colhead{K2ID \tablenotemark{d}} & \colhead{flag\tablenotemark{a}} 
}
\startdata
UCAC4-555-033290 & 101.764743 & 20.947725 & 11.705 & 11.192 & 11.067 & 11.927 & 11.661 & 11.756  &11.696 &11.726& 41.098906 &885.251787 &24 &202071861 &AAeeeAAAe \\
UCAC4-555-033327 & 101.789413 & 20.952493 & 14.487 & 12.700 & 12.157 & 15.417 & 14.481 & 14.920  &14.144 &13.857& 42.030470 &863.957681 &24 &202071849 &AAeeeAAAe \\
UCAC4-555-033328 & 101.789718 & 20.941804 & 15.207 & 14.082 & 13.808 & 15.967 & 15.332 & 15.603  &15.144 &14.984& 32.436217 &865.242115 &24 &202071849 &eeeeeeeee \\
UCAC4-555-033330 & 101.790216 & 20.942716 & 16.218 & 14.420 & 13.828 & 17.923 & 16.773 & 17.021  &16.130 &15.738& 33.183423 &864.694662 &24 &202071849 &eeeeeeeee \\
UCAC4-555-033335 & 101.795353 & 20.945481 & 11.554 & 11.266 & 11.224 & 11.589 & 11.466 & 11.478  &11.559 &11.677& 34.961871 &860.006732 &24 &202071849 &AAeeeAAAe \\
UCAC4-555-033336 & 101.795597 & 20.950077 & 16.008 & 14.560 & 14.051 & 17.444 & 16.468 & 16.771  &16.030 &15.720& 39.036805 &859.140643 &24 &202071849 &eeeeeeeee \\
UCAC4-555-033338 & 101.799223 & 20.951700 & 15.611 & 14.331 & 14.073 & 15.935 & 15.416 & 15.625  &15.249 &15.125& 39.999106 &855.879192 &24 &202071849 &eeeeeeeee \\
UCAC4-555-033418 & 101.876901 & 20.980552 & 14.091 & 13.383 & 13.252 & 14.427 & 14.080 & 14.193  &14.071 &14.044& 55.344496 &786.888096 &24 &202068459 &AAeeeAAAe \\
UCAC4-555-033420 & 101.876908 & 20.972664 & 12.859 & 11.979 & 11.702 & 13.485 & 12.948 & 13.181  &12.839 &12.712& 48.289514 &788.011635 &24 &202068459 &AAeeeAAAe \\
UCAC4-555-033425 & 101.881285 & 20.975126 & 16.224 & 13.317 & 12.709 & 17.317 & 16.082 & 16.193  &15.201 &14.762& 49.903488 &784.005099 &24 &202068459 &eeeeeeeee \\
... 

\enddata
\tablenotetext{a}{The photometry flag indicator: A-APASS photometry; T-Tycho photometry; e-estimated with 2MASS photometry}
\tablenotetext{b}{the $X$ and $Y$ coordinate of star at time 0, time 0 is defined as cadence 1 (BJD-2455895.528) for K2.0, 
and cadence 102 (BJD-2455975.178) for K2.1}
\tablenotetext{c}{the channel number of which the star is observed}
\tablenotetext{d}{the given K2 ID of the star}
\end{deluxetable*}
\end{turnpage}
%\end{landscape}
\clearpage

\begin{deluxetable}{ccc}
\tablewidth{0pc}
\tablecaption{Light Curve Segments used in EPD\label{table:seg}}
\tablehead{
\colhead{Segment No} & \colhead{start Cadence} & \colhead{end Cadence}
}
\startdata
1 & 0 & 454 \\
2 & 455 & 1005 \\
3 & 1006 & 1989 \\
4 & 2050 & 2314 \\
5 & 2315 & 2997 \\
6 & 2998 & 4020 \\
\enddata
\end{deluxetable}

\global\pdfpageattr\expandafter{\the\pdfpageattr/Rotate 90}

\end{document}